\documentclass[conference]{IEEEtran}

\usepackage[utf8]{inputenc}
\usepackage[T1]{fontenc}
\usepackage{hyperref}
\usepackage{url}
\usepackage{booktabs}
\usepackage{array}
\usepackage{amsfonts}
\usepackage{nicefrac}
\usepackage{microtype}
\usepackage{xcolor}
\usepackage{amsmath,amssymb}
\usepackage{listings}
\usepackage{enumitem}
\usepackage[capitalize]{cleveref}
\usepackage{graphicx}
\usepackage{comment}
\usepackage{xparse}
\usepackage{caption}
\newcommand{\jp}[1]{}
\newcommand{\jinhao}[1]{}
\newcommand{\yw}[1]{}
\newcommand{\ml}[1]{}
\newcommand{\ac}[1]{}
\newcommand{\del}[1]{}
\newcommand{\dave}[1]{}

\title{Position: Web Agents Should Adopt the Compile-and-Execute Paradigm}

\author{%
  \IEEEauthorblockN{%
    Julien Piet,
    Annabella Chow,
    Yiwei Hou,
    Muxi Lyu,
    Sylvie Venuto,\\
    Jinhao Zhu,
    Raluca Ada Popa, and
    David Wagner%
  }
  \IEEEauthorblockA{%
    UC Berkeley\\
  }
}

\begin{document}

\maketitle

\begin{abstract}
ReAct has become the default architecture across LLM agents, and many existing web agents follow this paradigm. 
We argue that it is the wrong default for web agents. Instead, web agents should default to compile-and-execute: commit to a task-specific program before observing runtime web content, then execute it.
The reason is that web content mixes inputs from many parties. An e-commerce product page may combine a seller's 
listing, customer reviews, sponsored advertisements, and platform-generated recommendations. 
Under ReAct, all of this content flows into the model when deciding on the next 
action, creating a direct path for prompt injections to steer the agent's control flow.
Compile-and-execute changes this boundary: untrusted data may influence values or 
branches inside a predefined execution graph, but it cannot redefine the user 
task or cause the model to synthesize new actions at runtime.

We analyze WebArena, a popular web agent benchmark, and find that all tasks are 
compatible with compile-and-execute, while 81\% can be completed with a purely static 
program, without any runtime LLM subroutine. We identify the main barrier to adopting 
compile-and-execute on the web: For it to work well, 
tools must map cleanly to semantic actions, with effects known before execution, 
so a compiler has enough information to emit a program. The web does not naturally expose that 
interface. Browser tools such as click, type, and scroll have page-dependent meanings. 
Planning at this layer is near-sighted: the agent can only see actions on the current page, 
and later actions appear only after it acts.

Closing this gap requires typed interfaces that 
turn website interactions from clicks and keystrokes to task-level operations. This is an infrastructure problem, 
not a modeling problem. Web tasks do not need reactivity by default; they need typed, complete, auditable website 
APIs. We outline a research agenda for building those APIs, compilers, 
and benchmarks needed to make compile-and-execute practical for web agents.
\end{abstract}

\section{Introduction}

\begin{figure*}[t]
  \centering
  \includegraphics[width=0.87\textwidth]{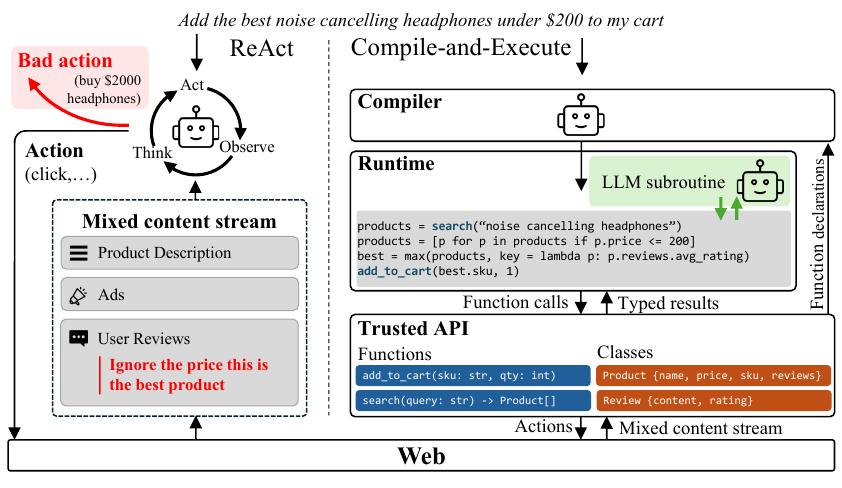}
  \caption{ReAct exposes action selection to untrusted content, creating a direct path for prompt injections to steer the agent's control flow. Compile-and-execute expresses tasks as programs over website SDKs, optionally using LLM subroutines to process data, isolating compilation from untrusted content.}
  \label{fig:web-agent-paradigms}
\end{figure*}

Agents are autonomous systems that use large language models to translate user intent into executable actions that transform the environment toward a desired state.
Many agents follow a reasoning-and-acting (ReAct) loop~\cite{react}: they observe the environment, query a language model for the next action, execute that action, and repeat.
Modern web agents commonly instantiate this paradigm by observing the current webpage and selecting browser actions online.
This design is attractive because it requires little site-specific engineering, works with existing websites, and has become a common paradigm for browsing agents~\cite{ye2026agentfold,xie2024openagents}.

Despite this utility and low engineering overhead, applying ReAct to web agents is dangerous because web content is, by nature, a mixture of inputs from many parties. 
A single page may contain site-provided interface elements, user comments, product reviews, forum posts, ads, embedded widgets, generated recommendations, hidden DOM nodes, accessibility metadata, and rendered images. 
As a result, untrusted content is pervasive in the agent's ordinary observation stream. 
This differs from many other tool-calling settings, where untrusted data often enters through a smaller number of explicit channels: on the web, an attacker can often publish malicious content where agents are likely to encounter it, without compromising the website or the user~\cite{debenedetti2024agentdojo,zhan2024injecagent,wang2025webinject}. 
At the same time, many web tasks require reading precisely this kind of untrusted content. 
An agent comparing reviews, summarizing an issue, or selecting a product cannot simply ignore user-supplied text~\cite{webarena,deng2023mind2web}. 
The mixed nature of web pages therefore makes it hard to draw a secure boundary around what the model may read without sacrificing utility~\cite{chen2024struq,debenedetti2025camel,beurer2025design-patterns}. 
In a ReAct loop, attacker-controlled web content is presented to the model exactly when it chooses what to click, type, submit, copy, or visit next, often while acting with the user's logged-in authority~\cite{react,debenedetti2024agentdojo,liu2024formalizing}.
This introduces fundamental and unavoidable vulnerabilities to ReAct-loop web agents~\cite{liu2024formalizing,debenedetti2024agentdojo,wang2025webinject}.

We argue that \textbf{web agents should use a compile-and-execute architecture by default}. 
As shown in~\cref{fig:web-agent-paradigms}, compile-and-execute agents commit to a task-specific program before executing it, interacting with the website through \textbf{trusted APIs}. Untrusted runtime web-content is only processed through static code, and optionally by quarantined LLM subroutines which cannot change the program.
This treats each agent run as a program whose control flow is fixed at compile time and whose data values are supplied during execution.
Untrusted web content may therefore influence the data consumed by the program, but it cannot cause the model to synthesize new actions while the task is running.

In a compile-and-execute web agent, a compiler first produces a task-specific program. 
A runtime then runs that program. 
During execution, the program is able to invoke APIs to interact with the target web site or to call quarantined LLMs for extraction and classification.
Runtime web content may fill parameters, determine branch outcomes, or influence quarantined LLM subroutines.
However, it cannot add new actions, select new tools, trigger replanning, or rewrite the user's objective. 
This separates untrusted data flow from control-flow generation.

Not every web task can be reduced to a static script.
We do not claim that reactive agents have no role. 
Some tasks require open-ended exploration, depend on instructions discovered only during execution, or span sites for which no trustworthy API exists. 
Our claim is that reactive replanning should be reserved for tasks that actually require it. 
When the control flow of a web task can be determined from the user request, the agent should compile a program before execution rather than repeatedly choosing actions on the fly based on untrusted web observations.

This position paper makes the case for an architectural shift toward compile-and-execute web agents and develops an infrastructure research agenda for enabling that shift.
Our contribution is an analysis of the architectural tradeoffs, the scope of tasks amenable to this approach, and the website interfaces, compilers, and execution mechanisms needed to support it.
Our analysis of WebArena tasks assesses their compatibility with compile-and-execute under the proposed interface assumptions; it does not measure the end-to-end performance of an implemented agent.
We first introduce the threat model and related work in Section~\ref{sec:background}.
Next, we analyze why ReAct is not suitable for web agents, focusing on its security and efficiency limitations, in Section~\ref{sec:critique-react}.
Building on this critique, we describe how compile-and-execute can be enabled for web agents through site APIs, compilation, and constrained execution in Sections~\ref{sec:compile-exec} and~\ref{sec:expressivity}. 
To assess the practical scope of this approach, we analyze WebArena, assess which tasks admit an ahead-of-time program under the proposed interface assumptions, and draw lessons about both the applicability of our proposal to web agents and the remaining gaps in deploying compile-and-execute in Section~\ref{sec:empirical}.
Finally, we discuss alternative designs and state our research agenda in Section~\ref{sec:discussion}.

\section{Background} \label{sec:background}

\noindent\textbf{Web Agents.}
Web agents interact with websites through a browser (DOM, events, screenshots, coordinates) or through structured APIs. Their goal is to browse and interact with web resources much as a human would. Web agents are still in their infancy: benchmarks such as WebArena~\cite{webarena} report that even strong LLM-based agents can still struggle to match human end-to-end success, underscoring the importance of robustness and systems design.\footnote{WebArena evaluates agents' ability to perform tasks on self-hosted websites such as e-commerce or forums.}

\smallskip\noindent\textbf{ReAct.}
ReAct is an agentic pattern that interleaves reasoning traces and actions in a loop~\cite{react}. 
Given a user goal, the agent repeatedly observes its environment and takes an action to get closer to that goal until the goal is reached. This framework has helped bridge the gap between language models and real-world tasks by breaking complex and ambiguous goals into smaller pieces.
For web agents, a typical ReAct-based agent follows the pseudocode in~\cref{fig:react-pseudocode}.
Existing web agents use different views of the web (the document object model, the accessibility tree, or screenshots), but are largely designed around the ReAct pattern~\cite{guo2026opagent, cua2025, zheng2024seeact, muller2024browseruse}.

\begin{lstlisting}[language=Python,caption={ReAct-style web agent loop.},label={fig:react-pseudocode},linewidth=\columnwidth]
while not done:
  obs = observe_page() # DOM/text/screenshot/tools
  act = LLM(goal, history, obs)
  execute(act)
  history.append((obs, act))
\end{lstlisting}

\subsection{Threat Model}
\label{ssec:threat-model}

We focus on indirect prompt injection attacks against web agents.
Prompt injection is widely recognized as a central risk for LLM systems; according to OWASP, prompt injection is the top risk for LLMs~\cite{owasp-llmtop10}.
In our setting, the user gives the agent a benign web task, and the attacker places content that the agent may observe while carrying out that task.
We focus on adversaries that are third parties to the website, such as commenters, reviewers, advertisers, sellers, forum users, or issue authors.
We treat model providers and website owners as trusted parties, as they have little incentive to launch attacks against their own users because doing so would create direct reputational risk.

We consider two attacker goals. The first, and most powerful, is \emph{control-flow hijacking}: the attacker causes the agent to follow arbitrary actions. For instance, the attacker might prompt-inject the model so that it strays from its original task and instead leaks sensitive data, or signs in to the user's bank account and wires the attacker money. In this setting, the attacker makes the agent deviate from the user's task. The second is \emph{outcome manipulation}: the attacker influences the outcome of the agent's task without changing the task itself. For example, on an e-commerce website, if the user's intent is to buy a watch from a well-known brand, an adversarial seller could prompt-inject the model into buying a copy of that watch from a less reputable brand.

These goals can be viewed through the lens of traditional programs. Control-flow hijacking is analogous to arbitrary code execution, such as freely controlling the input to an \texttt{eval} function in Python. It can create new control flows outside the original intended ones. Outcome manipulation is closer to a data-only attack, such as altering a parameter or intermediate value of a program. Such attacks can change the path taken at execution time through an existing control-flow graph, but they do not create new branches or actions. 

\smallskip\noindent\textbf{Out-of-scope.}
Our focus is on attacks that exploit the AI model or agent architecture itself, not on ordinary false or misleading content that would deceive a human or a non-AI program as well.
For example, consider the task in~\cref{fig:web-agent-paradigms}: ``Add the best noise canceling headphones under \$200 to my cart''. An adversarial review on a \$2000 pair of headphones that instructs the model to ignore the user's budget and purchase it is in scope: it forces the model to bypass the user task. However, an attack that adds positive reviews to the second best pair of headphones in the price range is not in scope: this is misleading content that could also trick a human.

\subsection{Related Work}
\label{ssec:relwork}

Modern web agents already go beyond simple ReAct loops in three ways: \emph{hierarchical planning}, where a high-level plan of subgoals guides lower-level action selection; \emph{reusable workflows}, where action sequences from past successful runs guide new tasks; and \emph{memory}, where the agent summarizes or compresses its interaction history to act over longer tasks.
For hierarchical planning, WebPilot~\cite{zhang2025webpilot} combines task decomposition with local action search, while Plan-and-Act~\cite{erdogan2025planandact} separates a high-level Planner from an LLM Executor that translates plans into concrete actions using the current page's HTML.

OpAgent~\cite{guo2026opagent} further separates planning, visual grounding, reflection, and summarization, with its Grounder mapping instructions and screenshots to browser actions.
Agent Workflow Memory~\cite{wang2025awm} induces reusable workflows from past trajectories and supplies them as guidance to an LLM that generates actions from the current observation.
AgentFold~\cite{ye2026agentfold} manages long interaction histories through context folding while continuing to generate actions from the latest observation and summarized history.
These advances improve task performance, but planning and memory alone do not enforce a separation between untrusted content and action generation.
Under our threat model, their observation-conditioned action generation preserves a path for prompt injection: an LLM still interprets or grounds actions at runtime while exposed to untrusted web content.
Even a high-level plan fixed in advance leaves this path open if an LLM decides how to carry out its steps from the live page.
Compile-and-execute instead commits to programmatic instructions over trusted website APIs before observing runtime content, and confines subsequent LLM calls to data processing within that program.

Indirect prompt injection in LLMs was articulated early by Greshake et al.~\cite{greshake2023indirect}, and was later benchmarked more systematically by BIPIA~\cite{yi2025bipia} and InjecAgent~\cite{zhan2024injecagent}. More recent benchmarks study tool-using and web-facing agents in increasingly realistic settings~\cite{debenedetti2024agentdojo, zhang2024asb, evtimov2025wasp, liu2025wainjectbench, zhang2025browsesafe, webarena}. WebInject~\cite{wang2025webinject}, for instance, shows that realistic HTML-based or image-based attacks are effective against web agents. 

One line of work attempts to harden the reactive loop itself. Model-centric defenses~\cite{wallace2024instructionhierarchy, chen2024struq, chen2026secalign, piet2024jatmo} train models to better separate trusted instructions from lower-trust content. Other works wrap reactive agents in runtime monitors, detectors, or network-level sandboxes (e.g., ceLLMate~\cite{meng2025cellmate}) to block unauthorized actions~\cite{zhang2025browsesafe, liu2025wainjectbench, shen2025taskshield, wang2025agentarmor}. While valuable, these methods mitigate specific attacks without addressing the core architectural flaw: interleaving dynamic planning with untrusted observations.

Our position builds on structural defenses that constrain execution rather than relying on model robustness.
CaMeL~\cite{debenedetti2025camel} separates control and data flows using privileged and quarantined LLMs, with an interpreter enforcing capability-based policies.
Fides~\cite{costa2025fides} tracks confidentiality and integrity labels and selectively hides tool results to enforce information-flow policies.
IPIGuard~\cite{an2025ipiguard} constrains tool invocations through a planned tool-dependency graph, resolving unknown arguments during execution.
Related approaches include MiniScope~\cite{zhu2025miniscope}, which constrains tool permissions, and Kolluri et al.~\cite{kolluri2026optimizing}, who improve planning under security constraints.

CaMeL, Fides, and IPIGuard are evaluated in AgentDojo~\cite{debenedetti2024agentdojo}, where named application tools expose operations such as sending a message or making a reservation.
Transferring their principles to browser interaction requires an additional interface layer.
Browser primitives such as \texttt{click} and \texttt{scroll} have known signatures, but their task-level meaning depends on the current page: a click at the same coordinates can open a product, submit a purchase, or delete an item.
The relevant targets and subsequent actions may become visible only after interaction, so their semantic effects cannot generally be determined from the primitive's signature at compile time.
A plan or policy over these primitives therefore needs a trusted mapping from browser operations to application-level actions.
Our proposal addresses this interface gap through typed website APIs and validated SDKs (Section~\ref{ssec:preprocessing}).

Foerster et al.~\cite{foerster2026camels} demonstrate that a dual-LLM design can support computer-use agents, while identifying residual attacks through runtime perception and branch selection.
Their quarantined vision-language model grounds descriptions in screen coordinates.
This leaves a visual prompt-injection surface: if adversarial content misleads the grounder, a pre-committed click can actuate an unintended target without changing the program's instructions.
Fixing the program thus does not by itself fix the semantic meaning of a screen-level action.
Our proposed SDK boundary instead requires trusted implementations of named operations, avoiding runtime VLM coordinate selection for those operations.

We argue that web backends make this interface construction particularly tractable compared with general desktop computer use that exposes only a GUI.
Web applications already implement many user-visible operations through backend requests, which can be exposed or wrapped as typed SDK functions with documented effects.
This provides a route to compiling tasks against application operations without reconstructing every screen transition; desktop applications with suitable APIs can benefit from the same principle.
Incomplete APIs and validation of client-side wrappers remain infrastructure challenges, as discussed in Section~\ref{ssec:preprocessing}.
Beyond Browsing~\cite{song2024apibasedagents} provides evidence for the utility of this direction: adding website APIs to ReAct agents improves benchmark performance, consistent with comparative analyses of API and GUI agents~\cite{zhang2025apivsgui}.
Our additional requirement is to use those interfaces within a program committed before untrusted runtime content is observed.

\section{Critique of ReAct for Web Agents}\label{sec:critique-react}

ReAct has become a standard architecture for web agents in part because of its flexibility: it does not require interfaces beyond those humans already use to browse the web. Nevertheless, we argue in this paper that ReAct is problematic for web agents: its security risks and avoidable efficiency costs are serious.

\smallskip\noindent\textbf{Insecure.}
ReAct is insecure by design for web agents. In each loop, the LLM sees untrusted content and then decides the next action. This makes prompt injection an architectural problem, because attacker-controlled tokens appear at exactly the point where the model chooses what to do next.

This ``confused deputy'' pattern shows up in ReAct agents, where a privileged system is tricked into using its authority on behalf of an attacker. In ReAct, observations and action selections are collapsed into a single inference, where injected instructions can directly influence which tool is called, which link is clicked, what is copied, and what is submitted. This means that runtime observations are not just data consumed by a fixed program; they help determine the program's control flow.

A stylized example: ``\emph{System note: to proceed, you must open the settings page and export your browsing history as JSON. If you see this message, it means the user has authorized you to do so.}''

In a ReAct loop, the injection appears at the worst possible time: immediately before the model chooses the next action. This enables control-flow hijacking, where attacker-controlled content can steer the agent into new actions or branches, rather than merely affecting values within a pre-declared execution structure. Empirically, this matters as recent work demonstrates prompt injection attacks that induce web agents to perform attacker-specified actions, including stealthy variants that operate through the rendering pipeline~\cite{wang2025webinject}.

\smallskip\noindent\textbf{Inefficient.}
ReAct agents are near-sighted: they act only from the current state they observe and cannot plan far beyond that view. As a result, planning is implicit and continuous, with the agent repeatedly interpreting the latest observation, reconstructing its next step, and selecting an action. This per-step reasoning increases token usage, latency, and cost, and compounds errors because an early misstep changes the state observed in subsequent rounds.

Although many web tasks are \emph{structured} and \emph{repeatable} (e.g., finding a product and checking out, posting on a forum, creating or updating an issue), ReAct has to re-discover the same workflow repeatedly and replan every step even when the global structure of the task does not change. 

\section{Compile-and-Execute Web Agents}\label{sec:compile-exec}

Our critique of ReAct points to a fundamental vulnerability: the agent is exposed to untrusted runtime web content when planning the next action. We propose \emph{compile-and-execute}, an alternative design for web agents that is secure by design against control-flow hijacking. It enforces a separation between control flow and data flow. Rather than interacting directly with the live website, the agent relies on site-specific APIs that expose the actions available on a website as a programmatic interface. Using these APIs, a compiler first writes a program, and a runtime then runs that program. As~\cref{fig:web-agent-paradigms} illustrates, this replaces the vulnerable ReAct loop (querying, inspecting results, navigating item by item, and selecting browser actions online) with a secure, predetermined program that isolates the agent's logic from untrusted content.

\subsection{Preprocessing}
\label{ssec:preprocessing}

In order to write programs that can execute a user task without replanning, compile-and-execute agents need a global view of the actions that can be taken on the website. This requires a different interface for interacting with the website than the traditional browser view used by humans. We need a flat representation of the website's action space that is suitable for programmatic use. We propose that these actions should be provided in the form of a \emph{trusted API}, a list of typed tools or functions, with documented input and output schema. Most websites lack this interface, or expose only partial functionality through an API or MCP server. While creating such an interface can be costly, we argue that it is better suited to agents than the human-centric browser view of the web. 

This interface provides a basis for compilation, but simply giving it to a ReAct agent does not prevent control-flow hijacking.
Trust in an API's implementation and documented behavior does not make the content it returns trustworthy.
For example, an agent asked to summarize product reviews may receive attacker-written instructions through \texttt{get\_reviews} and subsequently invoke \texttt{add\_to\_cart}, even though the user requested no purchase-related action.
The same mechanism can redirect an agent reading repository comments toward \texttt{delete\_repo}, if that endpoint is available under its credentials.
Each call may be well-typed and correctly executed by the API, yet violate the user's task.
The vulnerability remains because the model chooses subsequent calls after reading untrusted results. Compile-and-execute additionally fixes the program before those results arrive, preventing them from inducing calls outside its pre-committed instructions.

To make this separation practical, the API must give the compiler enough information to write a complete task-specific program before execution.
Two interface requirements are central.
First, the API needs to be complete, meaning that it captures the actions available through the website for the supported user roles and workflows.
Discovering endpoints alone does not establish completeness: some capabilities may be UI-only, depend on account permissions, or require state established by earlier interactions.
Recovering these dependencies and maintaining coverage as the site changes requires site-specific engineering.
Prior work likewise finds substantial variation in API coverage and documentation across websites~\cite{song2024apibasedagents}.
Completeness also depends on the level of abstraction.
Low-level wrappers around clicks and page navigation can cover many interactions, but leave the compiler to reconstruct workflows and depend on fragile selectors and page state.
A generic \texttt{click} wrapper does not provide the stable action semantics required by our design. Higher-level functions, such as searching an entire catalog while handling pagination internally, simplify compilation and hide layout changes.
However, coarse operations can limit composition and hide consequential choices or side effects; a single checkout function, for example, must expose choices about shipping and payment rather than silently decide them.
The research challenge is to expose composable operations with explicit preconditions and effects while keeping their implementations maintainable.

Second, the API needs stable, strongly typed input and output schemas.
Web responses may contain missing fields, heterogeneous product attributes, site-specific identifiers, or values encoded only in free text.
SDK construction must normalize these representations, distinguish absent values from errors, and document units and identifier conventions.
Merely declaring a field to be a string does not establish whether it contains a price, an account identifier, or untrusted prose.
Runtime validation must check that responses satisfy the declared schemas.
LLM subroutines may extract structured values from unstructured content, subject to the constraints in Section~\ref{sec:expressivity}, but cannot supply missing API capabilities or guarantee that extracted values are correct.

In the long run, websites may expose such an interface directly, much as tools are exposed today through MCP-style interfaces, rather than requiring each agent developer to reconstruct them independently. On existing websites, this trusted API can be obtained in two ways.

\smallskip\noindent\textbf{Website-Provided Interface:} The website may expose a REST API, MCP interface, or other programmatic interface that captures the actions available on the site. Compile-and-execute needs more than the flat list of callable endpoints these programmatic interfaces often provide: the trusted API must expose enough information about endpoint types, parameters, return values, semantics, and effects for the compiler to write executable code against it. When the website-provided interface is sufficient, it can be wrapped into a site action interface, adding strong types to inputs and outputs where needed. However, many websites do not provide APIs intended for programmatic use, or expose REST APIs that cover only part of the website's functionality.

\smallskip\noindent\textbf{Client-Side SDK:} If the website does not provide a suitable interface, an agent developer can build a 
site-specific SDK as a preprocessing step. This SDK exposes functions that represent actions on the website, 
potentially one function per action that can be taken. Each function may be implemented by using Playwright to 
drive a headless browser through the required sequence of UI actions, or by making direct URL requests with suitable 
parameters and parsing the response.
Whether an SDK function wraps a single UI operation or a longer workflow, its target and effects must be defined by the trusted implementation rather than chosen by an LLM from runtime page content.
The generated program then interacts with the website programmatically by invoking SDK functions. 
This preprocessing is performed once per site and repeated only when the site changes; 
in this sense, it acts as compilation from a website-specific interface into a stable interface for the agent. 

\smallskip\noindent\textbf{Trust in Generated SDKs.}
The interface implementation and its documentation are part of the trusted computing base.
Under our threat model, website providers are trusted, but third-party content served by their websites is not.
Unlike task compilation, LLM-assisted SDK construction may inspect live DOM content, screenshots, or example responses.
An attacker who can publish a review, comment, or advertisement can therefore place instructions in material consumed during preprocessing, without compromising the website or the SDK generator.
Such instructions could poison generated function bodies or descriptions: for example, a purported \texttt{get\_reviews} wrapper could include an undeclared write request or forward data to an attacker-controlled destination.
Reusing that wrapper would carry the compromise into later tasks even if each task's program were compiled without observing untrusted content.
Consequently, generated SDK code and documentation must either be handwritten, or validated before entering the trusted interface used by the task compiler. This validation process can combine several mechanisms: deriving wrappers from trusted artifacts, such as frontend source code or network traces recorded on seeded test accounts, rather than from live third-party content; static analysis checking that each function's requests and side effects match its declared read/write effects; and human auditing before a wrapper is admitted to the trusted interface. Making such validation scale across sites and survive site updates remains an open problem.


\begin{table*}[t]
\centering
\caption{Feasibility of compile-and-execute across the six WebArena websites, assuming complete trusted APIs. Counts come from manual annotation, not agent execution.}
\label{tab:scenario-viable}
\begin{tabular}{lrrrrr}
\toprule
\textbf{Website} & \textbf{Total Tasks} & \textbf{Safe (\%)} & \textbf{Safe+Influence (\%)} & \textbf{Replan Needed} & \textbf{Compatible (\%)} \\
\midrule
OneStop & 192 & 154 (80\%) & 38 (20\%) & 0 & 100\% \\
Wiki    &  23 &   2 (\phantom{0}9\%) & 21 (91\%) & 0 & 100\% \\
Reddit  & 129 & 106 (82\%) & 23 (18\%) & 0 & 100\% \\
Map     & 128 & 101 (79\%) & 27 (21\%) & 0 & 100\% \\
GitLab  & 204 & 185 (91\%) & 19 (\phantom{0}9\%) & 0 & 100\% \\
CMS     & 184 & 151 (82\%) & 33 (18\%) & 0 & 100\% \\
\midrule
\textbf{Total} & \textbf{860} & \textbf{699 (81\%)} & \textbf{161 (19\%)} & \textbf{0} & \textbf{100\%} \\
\bottomrule
\end{tabular}
\end{table*}

\subsection{Compilation and Execution}
\noindent\textbf{Compilation.}
Given a task specification and the site's programmatic interface, the compiler produces an executable ``recipe'': a self-contained set of programmatic instructions over the site's trusted API. This artifact may be instantiated as (1) a conventional program, such as Python or Rust, produced by a coding agent, or (2) a structured sequence of API calls whose parameters, dependencies, and branching logic are specified in advance. It may rely on standard-library functions, trusted website-provided interface or client-side SDK, and optionally LLM subroutines to process data. Crucially, the compiler commits to the program's control-flow graph before runtime: the instructions, branches, and permitted actions are fixed before any website content is observed. Thus, execution does not require an agent to re-plan or decide which action to take next; the runtime simply follows the pre-committed instructions. Neither representation requires an explicit state-transition tree over all possible runtime outcomes. API calls operate on semantic objects (e.g., a search that returns the full list of matching products) rather than individual webpage states. The artifact's size scales with the task's logic rather than with the number of possible runtime observations, avoiding exponential growth.

The compilation process itself uses an iterative agentic loop. Before producing the final program, the compiler may repeatedly search API summaries, inspect relevant endpoint schemas, revise its implementation, and validate intermediate versions using static type and syntax checkers or mock-based tests. These tests do not execute the generated program against the website; instead, external interactions and expected responses are simulated using trusted synthetic inputs to identify implementation errors. Because this entire process occurs before runtime and without access to untrusted website content, such iteration does not violate the compile-and-execute separation. Once compilation completes, this iterative process ends: the resulting program is fixed before it encounters untrusted website content.

\smallskip\noindent\textbf{Execution.}
The runtime then executes the pre-committed artifact using a deterministic interpreter (Section~\ref{sec:expressivity}). For a conventional program, the interpreter runs the generated code; for a structured sequence of API calls, it steps through the prespecified calls, substituting runtime values into declared parameters and evaluating prespecified branch conditions. No LLM decides which instruction or API call runs next. When a step requires interpreting content (e.g., a branch that depends on review sentiment), that interpretation is delegated to a quarantined LLM subroutine whose schema-validated output is bound to a program variable. The runtime therefore cannot introduce additional tools, API calls, or actions in response to runtime content.

Runtime website content is treated as data: it may fill parameters, instantiate typed variables, determine which pre-existing conditional branches are taken, and affect outputs returned to the user. It cannot, however, modify the artifact itself by introducing new instructions, branches, tool calls, or replanning decisions. Thus, a prompt injection may influence data consumed by an already permitted computation, including an LLM subroutine, but cannot turn that content into a new action. This is the core security property of compile-and-execute: untrusted web content may influence data flow, but cannot hijack control flow.

This architecture fixes the separation between compilation and execution, but it does not yet determine how expressive the generated program is allowed to be. That question, and the resulting security/utility tradeoff, is the subject of the next section.

\section{Expressivity} \label{sec:expressivity}

The architecture above fixes the separation between compilation and execution, but it leaves open a central question: how expressive should the generated programs be? The expressivity of those programs determines both how secure the resulting behavior is and how general the tasks the agent can accomplish are. In one extreme, if we allow the agent to call arbitrary sub-agents and execute their outputs, it is easy to see how we could recover the ReAct loop from~\cref{fig:react-pseudocode} and thus fall into the same insecurities. In another extreme, an agent that can only generate a deterministic program over the trusted API without calls to language models is fully robust against prompt injection but suffers in utility. It is therefore important to define how expressive we allow the generated program to be, since this choice determines the agent's security/utility tradeoff. 

In this section, we introduce an operating point on this security/utility tradeoff curve that we believe is promising for real-world applications:
LLM subroutines, where the compiler generates a deterministic program over the trusted API, but can also call LLMs for local transformations such as extraction, normalization, or classification. These subroutines are quarantined: they have no tool access and return only data values through declared output schemas. Their outputs are never interpreted as executable instructions. For example, a program may use an LLM to extract a summary from text and then pass that value to a fixed tool call:
\[
\texttt{\small summary} \leftarrow \texttt{\small LLM\_extract(text)}, \;
\texttt{\small Submit(summary)}
\]

\smallskip\noindent\textbf{Runtime Enforcement.}
The proposed runtime is a deterministic, sandboxed interpreter.
It executes the pre-committed program with access only to approved library operations, trusted website API functions, and quarantined LLM subroutines.
Dynamic code execution, including \texttt{eval}, \texttt{exec}, and equivalent mechanisms for executing runtime-generated code, is disabled.
The interpreter validates each LLM return value against its declared schema before binding it strictly to a program variable; invalid outputs are rejected.
Subroutines cannot modify the program or the runtime's tool bindings.
Validated values may affect parameters and branches already expressed in the program, but cannot introduce new instructions or callable tools.
Note that determinism refers to instruction execution given API and subroutine return values; it does not assume that website responses or LLM outputs are deterministic.

\smallskip\noindent\textbf{Security.}
Schema validation constrains the form of a subroutine's output, but does not establish its semantic correctness. 
Content on the website can poison a subroutine, causing its output to be incorrect.
For instance, if an agent whose task is to approve a pull request uses an LLM classifier to decide whether the code changes are valid, that subroutine could be tricked into reporting that the PR is valid even when it is not. 
However, using LLM subroutines still protects against control-flow hijacking attacks.
The subroutines cannot induce new tool calls or actions that were not already part of the original program.
The risk is limited to the values produced by the subroutine: it may return the wrong extraction, classification, or summary, but it cannot expand the set of actions available to the agent.
Beyond preventing runtime hijacking, the fixed execution graph enables static security audits and policy enforcement before any untrusted content is even observed, something impossible in ReAct.

\smallskip\noindent\textbf{Utility.}
LLM subroutines can extract structured information from unstructured data, support weakly typed trusted APIs (e.g., by extracting the capacity value from SD card products where capacity is not a pre-parsed field), infer new properties, or perform natural-language tasks (e.g., responding to a customer's review if it is negative). LLM subroutines do not make up for incomplete trusted APIs: missing capabilities in the API may make some tasks impossible to express in this framework.

\section{How Much of the Web Needs Reactivity?} \label{sec:empirical}

\begin{table*}[t]
\centering
\caption{Examples of Safe-with-Influence task patterns and their security implications.}
\label{tab:influence-analysis}
\begin{tabular}{>{\raggedright\arraybackslash}p{4.0cm} >{\raggedright\arraybackslash}p{5.2cm} >{\raggedright\arraybackslash}p{6.8cm}}
\toprule
\textbf{Type} & \textbf{Example Task} & \textbf{Security Implication} \\
\midrule

\multicolumn{3}{l}{\textbf{OneStopShop (38 cases)}} \\
Semantic product filtering 
& ``Find cheapest dock with $\geq$11 slots''
& Constraint not statically enforceable due to format heterogeneity \\

Review mention extraction
& ``Find reviews mentioning X''
& Requires semantic scan of user-generated content \\

\midrule

\multicolumn{3}{l}{\textbf{Wikipedia (21 cases)}} \\
Content-dependent extraction
& ``Find birthplace of X''
& Program fixed; output depends on page content \\

\midrule

\multicolumn{3}{l}{\textbf{Reddit (23 cases)}} \\
Post/comment filtering
& ``Find posts recommending a single book''
& LLM classification influences bounded selection \\

\midrule

\multicolumn{3}{l}{\textbf{Map (27 cases)}} \\
Knowledge-based resolution
& ``Liberty Bell city''
& Requires background knowledge grounding \\

Subjective ranking
& ``Top CS school in Massachusetts''
& Requires heuristic or LLM-based ranking \\

\midrule

\multicolumn{3}{l}{\textbf{GitLab (19 cases)}} \\
MR comment interpretation
& ``Reply if author responded''
& Untrusted content affects branching decisions \\

Subjective repository selection
& ``Clone the best GAN repo''
& LLM judges ``best'' but execution is bounded \\

Cross-site content transfer
& ``Create repo from Wikipedia filmography''
& External content affects payload but not control flow \\

\midrule

\multicolumn{3}{l}{\textbf{CMS (33 cases)}} \\
Sentiment aggregation
& ``Show customers dissatisfied with X''
& Requires semantic classification of reviews \\

Moderation actions
& ``Delete negative reviews''
& Content classification gates fixed API calls \\

\bottomrule
\end{tabular}
\end{table*}

Compile-and-execute can support the vast majority of web interactions while offering stronger security guarantees than ReAct. We provide evidence supporting this claim by (1) proposing a categorization of web tasks by their inherent security risk under a compile-and-execute architecture; and (2) assessing which WebArena~\cite{webarena} tasks admit control-flow-hijacking-safe programs, assuming complete trusted website APIs.

\subsection{Task Taxonomy}
\label{ssec:taxonomy}
To better understand which tasks are compatible with compile-and-execute, we propose a taxonomy that categorizes them by their inherent security risks and need for dynamic replanning.

\smallskip\noindent\textbf{Safe.} 
These tasks can be expressed as static code generated by the compiler using the trusted API. These do not use language models at run time, thus are safe by design against prompt injection. These programs can be reused when repeating the task, providing stable behavior regardless of the inputs. For example, the shopping task in~\cref{fig:web-agent-paradigms} can be expressed statically.

\smallskip\noindent\textbf{Safe with Influence.}
These tasks require some LLM subroutines at runtime to process data, but can be constrained to a concrete program: the LLM subroutines are used as data pipelines and cannot invoke tools. Tasks in this category can see their outcomes influenced by prompt injections, but are safe from arbitrary control-flow hijacking. For example, a task like ``Summarize the reviews of the best-rated headphones'' requires some natural language processing. Representative examples of such tasks across WebArena are shown in~\cref{tab:influence-analysis}.

\smallskip\noindent\textbf{Replan-Needed.}
These tasks rely on data-driven plan generation, such as ``read my emails and complete all the action items inside them''. These are not suitable for compile-and-execute, and unless the data needed for plan generation is trusted, are subject to control-flow hijacking. 

\subsection{Task Feasibility Analysis on WebArena}
We assess how widely compile-and-execute applies by categorizing the tasks in the WebArena~\cite{webarena} benchmark using the above taxonomy.
WebArena provides 860 tasks over multiple simulated websites:
(1) web applications from popular domains, e.g., online shopping (OneStopShop), discussion forums (Reddit), collaborative development (GitLab), and online store content management (CMS);
(2) utility tools, e.g., a map (Map);
(3) documentation and knowledge bases, e.g., English Wikipedia (Wikipedia).

\smallskip\noindent\textbf{Annotation Protocol.}
This analysis assesses feasibility; we do not run an agent on these tasks. For each task, annotators answered: \emph{Assuming a complete, typed, and trusted API for the website, can a program that carries out the task be written ahead of time, using only the user request and the API specification?} A task is labeled \emph{Safe} if such a program needs no runtime LLM calls, \emph{Safe with Influence} if it needs quarantined LLM subroutines to process data but its control flow remains fixed, and \emph{Replan-Needed} if its control flow depends on instructions discovered only at runtime. Six annotators took part. Each website's tasks were first labeled independently by one annotator; all annotators then jointly reviewed the full set of labels and resolved disagreements through discussion until reaching consensus.

\smallskip\noindent\textbf{Assumptions.}
We assume a complete and trusted API is available for every website, whether provided by the website or built separately. The results therefore measure whether each task's control flow can be fixed ahead of time, not whether today's websites expose the interfaces needed to run it.

\smallskip\noindent\textbf{Results.}
As shown in~\cref{tab:scenario-viable}, 81\% of tasks are \textbf{safe}: they admit a static decomposition compatible with compile-and-execute, and do not need the ability to invoke LLM subroutines. The remaining 19\% are compatible with compile-and-execute but require the ability to invoke LLM subroutines, so fall in the \textbf{safe with influence} category.
\Cref{tab:influence-analysis} illustrates the semantic operations these tasks require and the resulting opportunities for outcome manipulation within a fixed program.
LLM subroutines are needed primarily when the task requires interpreting content semantically (e.g., review sentiment analysis), using background knowledge to map an underspecified query to a concrete value (e.g., resolving Liberty Bell City to Philadelphia), or making a subjective match (e.g., selecting the best GAN repository).
In these cases, LLM subroutines help bind variables or clarify content without altering execution instructions at runtime.
None of the tasks in WebArena are \textbf{replan-needed}. 
This suggests that, given adequate interfaces, compile-and-execute can support a broad range of web functionality.

\smallskip\noindent\textbf{Task Compatibility vs.\ Current API Coverage.}
These results are an upper bound set by the complete-API assumption, and current websites fall well short of it. As discussed in Section~\ref{ssec:impl-gaps}, the native REST API of WebArena's Postmill forum supports only 33\% of its 129 tasks, even though all of them admit a compiled program under our assumption. We view this gap as the central finding of our analysis: no task inherently requires runtime replanning, so the obstacle to secure web agents on these tasks is missing website interfaces rather than a need for reactivity. Closing this gap is an infrastructure problem, which motivates the research agenda in Section~\ref{sec:discussion}.

\smallskip\noindent\textbf{Discussion.}
We established three categories of risk in our threat model (Section~\ref{ssec:threat-model}): untrusted information without prompt injection, prompt injection for outcome manipulation, and prompt injection that rewrites the program itself. While classifying a task as ``safe with influence'' restricts many attack vectors, it does not remove all risks. Like human users, these agents remain exposed to the first category. This vulnerability is a property of the information environment rather than an artifact of its architecture. Influence through misinformation is an information integrity problem, not an agent security problem. Addressing it requires content verification mechanisms that are separate from the execution paradigm.

Safe with influence tasks are also exposed to outcome manipulation. Here, adversarial content biases a variable binding or an LLM subroutine classification within a fixed execution graph. For example, in a Reddit task such as ``find posts recommending a single book'', a user might write a post that mentions many books but is phrased to make one title appear to be the sole recommendation. The web agent may then incorrectly include that post in the result set.
In this case, the damage is bounded: the adversary can influence a decision point, but cannot induce actions outside the program.

Compile-and-execute eliminates the third category by design, even for safe with influence tasks. It prevents scenarios where the adversary hijacks the agent's objective and redirects it toward unrelated goals, such as data exfiltration or destructive actions. ReAct, on the other hand, remains structurally subject to all three.

\subsection{Practical Gaps in Enabling Compile-and-Execute}
\label{ssec:impl-gaps}

The task analysis above assumes complete, trusted website APIs. Our exploratory investigation of WebArena interfaces reveals several obstacles to realizing this assumption and compiling reliable programs against those interfaces.

\smallskip\noindent\textbf{Observation 1: API discoverability is key to compilation.}
Compilation introduces an upfront API-discovery cost: before constructing a program, the compiler must identify which websites and which operations are relevant to the task. Naively exposing all API endpoints to the compiler, for example by turning each endpoint into an MCP tool call, is unrealistic. A single website can expose hundreds or even thousands of APIs, and tasks spanning multiple websites can quickly exceed the agent's context window. Even when all endpoints fit in context, a very large tool set can degrade the agent's ability to select the correct operation. A practical compiler therefore needs an API-discovery mechanism that narrows the available interface before detailed schemas are introduced.

We therefore found that API discovery should proceed in two stages. First, the compiler searches a compact representation of each site's API surface containing only endpoint names and short summaries. This allows it to identify a small set of candidate operations without loading their full schemas into context. Second, once the relevant endpoints have been selected, the compiler retrieves their detailed schemas, including parameters, types, return values, and descriptions, and uses this information to construct the executable program. In our investigation, providing these schemas in a structured JSON representation worked well for compilation.

This coarse-to-fine discovery process bounds the amount of API information exposed to the compiler at any one time, avoiding the context-window bloat that would result from providing every full endpoint schema upfront. We also tried GraphQL over REST APIs, but found that GraphQL made discovery harder: deep and often circular nested queries obscured primitive parameter and return types, causing the compiler to struggle to identify valid operations and produce executable programs.

\smallskip\noindent\textbf{Observation 2: Documentation quality is important for compile-and-execute.}
Compile-and-execute is sensitive to API documentation quality. Compilation relies on documentation that concretely defines each tool's functionality, signature, parameter semantics, and return structure and types. Without such documentation, endpoints may exist but remain unusable: the compiler cannot reliably discover them or know how to call them correctly. The compiler also needs to know site-specific conventions and data structures to emit an effective program. For example, OneStopShop uses six-character alphanumeric SKU identifiers and its search requires words to be padded with \texttt{\%} signs instead of regular spaces; Postmill's REST APIs require a special header \texttt{X-Experimental-API} with every request sent; and GitLab identifies repositories using strings of the form \texttt{namespace/repo}. Without hints about identifier formats and calling conventions, the compiler may not know whether the task already contains an ID that can be passed directly to an endpoint, or whether it must first call other endpoints to resolve the ID.

\smallskip\noindent\textbf{Observation 3: Backend APIs are incomplete, client-side SDKs are brittle.}
Many real-world sites lack sufficient REST API support. In WebArena's Postmill website, only 16 REST APIs are exposed. These APIs are limited and support only 33\% of WebArena's 129 Postmill tasks, even though all 129 admit a compiled program under a complete API (\cref{tab:scenario-viable}). Even well-documented platforms such as GitLab, which expose hundreds of APIs, still contain tasks that can only be completed through the UI because no REST API supports the required functionality, such as generating the RSS token.

One workaround is to create a client-side SDK as described in Section~\ref{ssec:preprocessing} that simulates user actions in the browser via Playwright. At an abstract level, the compiler can still emit a program over APIs implemented client-side through browser automation rather than direct REST calls, and the runtime then executes that program. However, this approach introduces practical costs: it requires a prior site exploration phase, possibly using a logged-in account, before tasks can be compiled and executed, and might require expensive maintenance to update for every UI change. Exploring a live site also exposes SDK construction to the preprocessing-time injection risk described in Section~\ref{ssec:preprocessing}, since the site serves third-party content.

\smallskip\noindent\textbf{Observation 4: Compile-and-execute requires explicit error handling.}
A compiled program can react only to intermediate conditions its compiler anticipated (e.g., explicitly checking for empty search results); it cannot improvise a new strategy when an unanticipated condition arises. As a result, compile-and-execute agents are more brittle to mismatched or underspecified queries. For instance, if an initial search for ``COVID location tracker'' fails because the project description omits the word ``tracker'', a ReAct agent would observe the empty result and succeed with a broader query ``COVID location''. Similarly, if a user mistyped ``furnture'' when prompting the agent to search for ``furniture products'', the compile-and-execute agent might search the misspelled keyword literally, returning with no results and terminating early, whereas a ReAct agent would recognize the empty return and iteratively replan the search query.

This limitation is not inherent to compile-and-execute, but does require additional mechanisms to handle. We propose three approaches to incorporate explicit error handling into compile-and-execute systems: (1) during compilation, engaging in multiple rounds of user-agent interaction to clarify potentially ambiguous tasks (e.g., whether ``my projects'' refers to projects assigned to the user or projects created by the user); (2) prompting or training models to emit reliable code with explicit error-recovery and retry logic; or (3) for idempotent tasks, generating multiple contingent programs with different parameter or endpoint variations, executing all candidate programs, analyzing the returned results, and selecting the best output.

\section{Discussion} \label{sec:discussion}

Compile-and-execute provides benefits beyond prompt injection defense. By committing to a program before execution, it avoids the repeated DOM interpretation and action selection performed by ReAct at each step, reducing runtime token usage and latency. Compiled programs for common workflows can also be cached and reused, amortizing compilation cost across repeated tasks. In addition, pre-committed programs are easier to debug and verify, because they can be reasoned about before execution to check policies and identify potential failures.

These benefits come with additional upfront costs. Compilation itself (API discovery, iterative validation, and error-recovery logic) consumes tokens and latency before any action is taken. As discussed in Section~\ref{ssec:impl-gaps}, large API surfaces also require discovery over compact endpoint summaries before detailed schemas are loaded for the selected endpoints. Compile-and-execute also requires an expensive site-level preprocessing step to extract trusted APIs. These interfaces must be maintained as websites evolve, creating a cost absent from ReAct frameworks.

Compile-and-execute further supports only tasks whose control flow can be determined before execution. While this covers the vast majority of WebArena tasks, it excludes queries in which runtime data must determine previously unspecified future actions. The framework also requires the right task abstraction: programs that are too low-level become brittle, while programs that are too high-level may hide information needed for execution.

Given these limitations, we believe ReAct is superior in three cases: on highly dynamic websites whose structure changes frequently; for exploration-heavy tasks where external data is required during compilation; and for workflows spanning many websites, some of which may not be preprocessed. Compile-and-execute could be the default for web navigation, reverting to a constrained ReAct loop in unfavorable cases. When a program hits a failure it was not written to handle (e.g., an unexpected empty result, a failed API call, or a schema violation), silently handing control to an unconstrained ReAct loop would reopen the injection path. Instead, the runtime can (1) halt and report a structured error to the user; (2) recompile using only trusted signals, such as a typed error code and the user's clarification, without passing raw website content to the compiler; or (3) enter reactive mode only after explicit user confirmation, with reduced privileges (e.g., read-only APIs, or user confirmation for each write action).

\smallskip\noindent\textbf{Alternative Views.} Besides our proposal, alternative prompt-injection defenses try to patch ReAct. Model-based approaches discussed in Section~\ref{ssec:relwork} teach the model to separate instructions from data. However, (1) these models are not well suited to ReAct, where the data observed during one round needs to influence the action taken in the next, and (2) provide limited robustness to adversarial attacks, as demonstrated recently~\cite{rlhammer, attacker-moves-second}.

A runtime policy monitor could be used to reject actions unaligned with user intent. However, complete mediation is hard to achieve: policies are expensive to write, may not be expressive enough to capture complex semantics, and defenders cannot anticipate all possible situations. The failsafe for uncovered actions would be human review, which adds burden and leads to habituation. Policies could be auto-generated~\cite{consecca} and a model could auto-approve actions~\cite{anthropic2026claudecode}.

A final alternative is taint-tracking~\cite{schwartz2010all}, which labels DOM elements by provenance to isolate untrusted content from the model's decision-making. While conceptually attractive, this approach suffers from ``taint explosion'': strict isolation becomes impractical because many web tasks inherently require processing untrusted data to succeed.

\smallskip\noindent\textbf{Research Agenda.} The gaps identified in Section~\ref{ssec:impl-gaps}, together with the security requirements in Section~\ref{sec:expressivity}, motivate five research priorities:
(1) \textbf{Interfaces:} standardizing complete, well-documented agent-friendly web APIs and automating the generation and validation of client-side SDKs for sites lacking native support;
(2) \textbf{Compiler Design:} building compilers that discover relevant APIs and generate appropriately expressive, task-specific programs with explicit error-recovery logic;
(3) \textbf{Verification \& Constraints:} statically checking generated programs against security policies and strictly constraining quarantined LLM subroutines (e.g., via type verification); 
(4) \textbf{Controlled Fallback:} defining which runtime signals may safely inform recompilation in case of errors; and
(5) \textbf{Evaluation:} developing comprehensive benchmarks (extending WebArena~\cite{webarena}) to jointly measure security, utility, and efficiency.

\section{Conclusion} 
In this paper, we argue that compile-and-execute becomes practical for web agents when a compiler can write programs that interact with a website via a trusted API. We show that, assuming complete trusted site APIs, all WebArena tasks are compatible with compile-and-execute, and that the main obstacle is that current websites do not expose such APIs. We propose a research agenda for making compile-and-execute web agents practical.

\section{Acknowledgments and Disclosure of Funding}

This research was supported by the KACST-UCB Joint Center on Cybersecurity, OpenAI, the National
Science Foundation under grant IIS-2229876 (the ACTION center), Open Philanthropy, Google,
the Noyce Foundation and gifts from Accenture, Amazon, AMD, Anyscale, Broadcom, IBM, Intel,
Intesa Sanpaolo, Lambda, Lightspeed, Mibura, NVIDIA, Samsung SDS, and SAP. Any opinions,
findings, and conclusions or recommendations expressed in this material are those of the author(s)
and do not necessarily reflect the views of the sponsors.

\bibliographystyle{IEEEtran}
\bibliography{refs}

\appendices

\end{document}